\newif\ifOnline
\Onlinetrue

\newif\ifPP
\PPtrue

\documentclass{emulateapj}
\usepackage{amsmath}
\usepackage[varg]{txfonts}

\usepackage{url}
\usepackage{graphicx}
\usepackage{amssymb}

\ifOnline

  \usepackage[pdfusetitle,pdftex,breaklinks,backref,colorlinks=true]{hyperref}

  \let\url\relax
\fi

\ifPP
  \makeatletter
  \renewcommand{\fps@figure}{tp}
  \makeatother
  
  \slugcomment{To appear in Radio Science.}
\else
  
  \newcommand{\keywords}[1]{} 
\fi

\newcommand{\mathe}{\ensuremath{\mathrm{e}}}
\newcommand{\mi}{\ensuremath{\mathrm{i}}}
\newcommand{\dif}{\ensuremath{\mspace{1mu}\mathrm{d}\mspace{-1mu}}}

\newcommand{\jones}[2]{\ensuremath{\mathbf{\mathsf{#1}}_{#2}}}

\ifOnline
  \newcommand{\rampname}{colorscale}
  \newcommand{\negcolor}{blue}
  \newcommand{\poscolor}{red}
\else
  \newcommand{\rampname}{grayscale}
  \newcommand{\negcolor}{black}
  \newcommand{\poscolor}{white}
\fi

\begin{document}
 

\title{Correcting The Polarization Leakage Phases and Amplitudes Throughout the
  Primary Beam of an Interferometer}

%
%

\ifPP
  \author{R. I. Reid\altaffilmark{1}, A. D. Gray, T. L. Landecker, and
          A. G. Willis}
  \affil{Dominion Radio Astrophysical Observatory, Herzberg
    Institute of Astrophysics, National Research Council, P.O. Box 248,\\
    \phantom{$^2$}Penticton, BC, Canada, V2A 6J9.}
  \email{rreid@nrao.edu}

  \altaffiltext{1}{Now at the National Radio Astrophysical Observatory,
    520 Edgemont Rd., Charlottesville, VA, 22903, USA}
\else
  \authors{R. I. Reid, \altaffilmark{1, 2}
    A. D. Gray, \altaffilmark{1}
    T. L. Landecker, \altaffilmark{1}
    and A. G. Willis\altaffilmark{1}}

  \altaffiltext{1}{Dominion Radio Astrophysical Observatory, Herzberg
    Institute of Astrophysics, National Research Council, P.O. Box 248,
    Penticton, BC, Canada, V2A 6J9.}

  \altaffiltext{2}{National Radio Astrophysical Observatory,
    520 Edgemont Rd., Charlottesville, VA, 22903, USA}
\fi

\begin{abstract}
  Polarimetric observations are affected by leakage of unpolarized light into
  the polarization channels, in a way that varies with the angular position of
  the source relative to the optical axis.  The off-axis part of the leakage is
  often corrected by subtracting from each polarization image the product of
  the unpolarized map and a leakage map, but it is seldom realized that
  heterogeneities in the array shift the loci of the leaked radiation in a
  baseline-dependent fashion.  We present here a method to measure and
  remove the wide-field polarization leakage of a heterogeneous array.  The
  process also maps the complex voltage patterns of each antenna, which can be
  used to correct all Stokes parameters for imaging errors due to the primary
  beams.
\end{abstract}

\keywords{instrumentation: interferometers ---  instrumentation: polarimeters
  ---  techniques: interferometric ---  techniques: polarimetric}

\ifPP
\else
  \begin{article}
\fi

\section{Introduction}
\label{sec:intro}

The hardware typically used in radio telescopes has the great benefit of
observing the Stokes $Q$, $U$, and $V$ parameters simultaneously with Stokes
$I$, but always allows some mixing between the polarization channels, as in
Figs.~\ref{fig:intdiag}~and~\ref{fig:nolc}.  This ``leakage'' is particularly
troublesome when it goes from $I$ into $Q$, $U$, or $V$, since the polarized
signals are usually a small fraction of the total intensity, and therefore
easily swamped by similarly strong leakages from $I$.

\begin{figure}
  {\centering
  \plotone{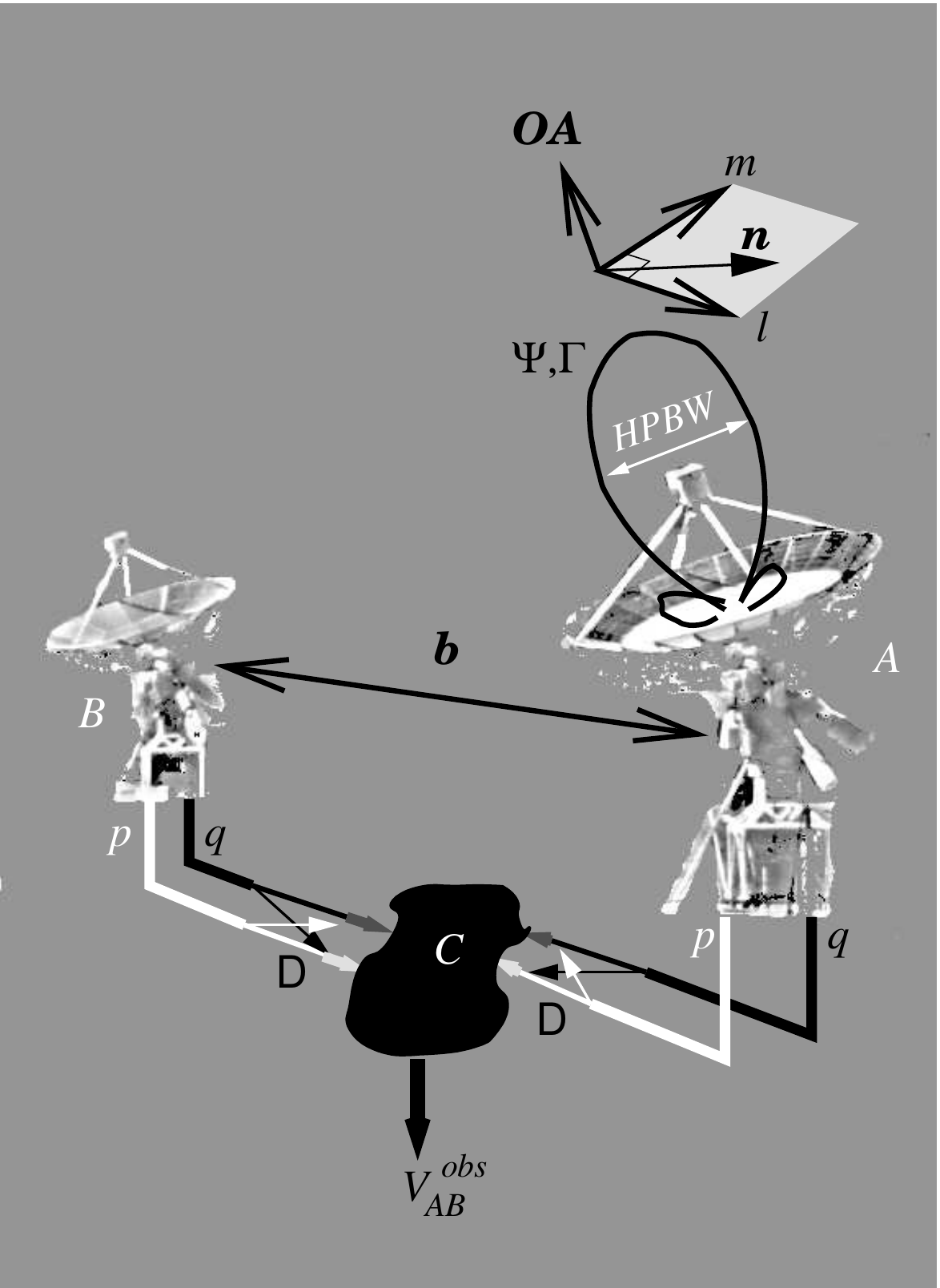}
  \caption{Conceptual diagram of polarization leakage in an interferometer.
    Each antenna measures two nominally orthogonal polarizations $p$ and $q$,
    but they are partially mixed before entering the correlator $C$.
  }
  \label{fig:intdiag}
  \begin{tabular}{r@{: }p{0.85\linewidth}}
    $\boldsymbol{b}$ & Baseline (separation) between antennas.\\
    $\boldsymbol{O\!A}$ & The optical axis (i.e.\ pointing direction).\\
    $(l, m)$ & Longitudinal and latitudinal offsets perpendicular to $\boldsymbol{OA}$.\\
    $\boldsymbol{n}$ & An arbitrary offset in $(l, m)$.\\
    \jones{\Gamma}{A} & Voltage pattern of antenna $A$, factored to exclude
    polarization leakage.\\
    HPBW & Half Power Beam Width.\\
    \jones{D}{B} & $\boldsymbol{n}$ independent factor of the polarization
      leakage of antenna $B$.\\
    \jones{\Psi}{} & $\boldsymbol{n}$ dependent
      factor of the polarization leakage.\\
    $V_{\mathrm{obs},AB}$ & Observed vector of visibilities in each polarization.
  \end{tabular}}
  Although the diagram places \jones{\Psi}{} and \jones{\Gamma}{} above the
  receiver and \jones{D}{} below, each includes effects from the feed,
  reflector surface, and receiver support struts.
\end{figure}

A radio interferometer uses two or more antennas to measure the amplitudes and
phases of the electric field impinging on their receivers.  The measurements
are stored as ``visibilities'', which are the correlations of the receiver
voltages.  Given some conditions which this article will assume to have been
met, the visibilities sample the Fourier transform of the sky multiplied by the
primary beam (directional sensitivity function) of the antennas
\citep{bib:siss_clark, bib:siss_thompson}.

Mathematically, the effect of a pair of antennas $A$ and $B$ on the
visibilities they observe, $V_{\mathrm{obs},AB}$, is conveniently expressed
using the Hamaker-Bregman-Sault \citep{bib:hbs1} formalism, where the four
polarizations are combined into a column vector.  The true visibilities are
multiplied on the left by a set of Jones matrices, each one the outer product
of Jones matrices for antennas $A$ and $B$, i.e.
\begin{equation*}  
  \jones{D}{AB} = \jones{D}{A} \otimes \jones{D}{B}^*
\end{equation*}
represents the on-axis mixing between the nominally orthogonal polarization
channels, often called the ``$D$ terms''.  The outer product of two matrices
$\mathbf{\mathsf{M}}$ and $\mathbf{\mathsf{N}}$, $\mathbf{\mathsf{M}} \otimes
\mathbf{\mathsf{N}}$, is formed by multiplying each entry of
$\mathbf{\mathsf{M}}$ with all of $\mathbf{\mathsf{N}}$, and is used, along
with a complex conjugation of the second factor, to bring together the elements
from each antenna in a correlation.  \citet{bib:hbs1} explain the algebraic
properties, including coordinate transformations, of Jones matrices and the
outer product in more detail.  As in \citet{bib:bhatnagar_evla100}, direction
dependent effects can also be included, but they must go \emph{inside} the
Fourier integral:
\ifPP
  \begin{gather}
    V_{AB}^\textrm{obs} = \jones{D}{AB}\!\int\! 
    \jones{\Psi}{\! AB}(\boldsymbol{n})
    \jones{\Gamma}{\! AB}(\boldsymbol{n})
    \mathbf{\mathsf{S}}I_S(\boldsymbol{n} + \boldsymbol{n}_c)\,
    \mathe^{\mi \boldsymbol{n} \cdot \boldsymbol{b}_{AB}}\dif n
    \label{eq:js}
  \end{gather}
\else
  \begin{align}
    V_{AB}^\textrm{obs} = \jones{D}{AB}\!\int\! &
    \jones{\Psi}{\! AB}(\boldsymbol{n})
    \jones{\Gamma}{\! AB}(\boldsymbol{n})\cdot\nonumber\\
      & \mathbf{\mathsf{S}}I_S(\boldsymbol{n} + \boldsymbol{n}_c)\,
      \mathe^{\mi \boldsymbol{n} \cdot \boldsymbol{b}_{AB}}\dif n
    \label{eq:js}
  \end{align}
\fi
where $\boldsymbol{n}$ is a direction on the sky relative to the ``phase
tracking center'', $\boldsymbol{n}_c$.  $\boldsymbol{n}_c$ is set
electronically, but usually it is chosen to coincide with the pointing
direction of the antennas.  $\mathbf{\mathsf{S}}$ is the Stokes matrix, which
transforms the sky's Stokes parameters, $I_S = (I, Q, U, V)$, into the
observational polarization basis, typically correlations of either circular or
linear polarizations.  \jones{\Psi}{AB} and \jones{\Gamma}{AB} are respectively
the wide-field leakage pattern and primary beam for the correlation of antennas
$A$ and $B$.  They are sometimes multiplied together to form a single Jones
matrix which is a generalization of the primary beam, but the magnitudes of the
effects are more easily assessed if they are kept separate.  With the
separation, \jones{\Gamma}{AB} is diagonal since it does not mix polarizations
in the observational basis, and the diagonal elements of \jones{\Psi}{AB} are
all one.

The on-axis portion of the leakage, \jones{D}{AB}, is dealt with by
standard polarimetric calibration techniques, but the leakage varies with
direction, growing worse toward the edges of the primary beam, as in
Fig.~\ref{fig:nolc}.  This paper is concerned with the wide-field
polarization leakage, \jones{\Psi}{AB}, and will assume that
$V_{\mathrm{obs},AB}$ has already been corrected by multiplication with
$\jones{D}{AB}^{-1}$.
 
\begin{figure*}
  \begin{minipage}[t]{1.0\columnwidth}
    \plotone{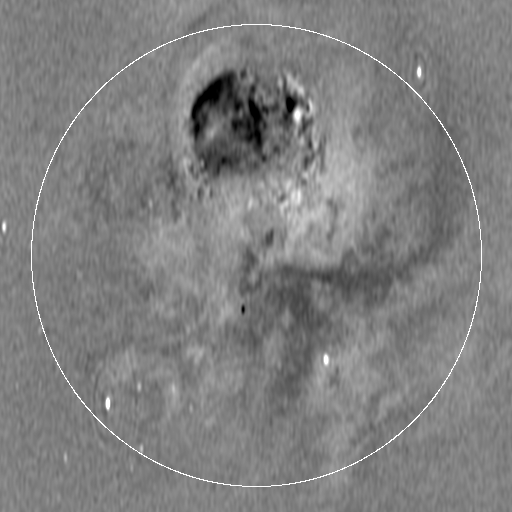}
    \caption{An example of leakage from Stokes $I$ into Stokes $U$ in a CGPS
      field containing the supernova remnant IC443.  The thin circle is the
      75\arcmin\ radius (24\% power) cutoff of the usable part of the beam in
      polarization.  The image has not been corrected for the sensitivity
      dropoff of the primary beam, and only includes ST data.  Note that it
      has been CLEANed, so the arcs are mostly leakage.  The grayscale goes
      from -5 (black) to 5 (white) mJy/beam. \label{fig:nolc}}
  \end{minipage}\hfill\begin{minipage}[t]{1.0\columnwidth}
  \plotone{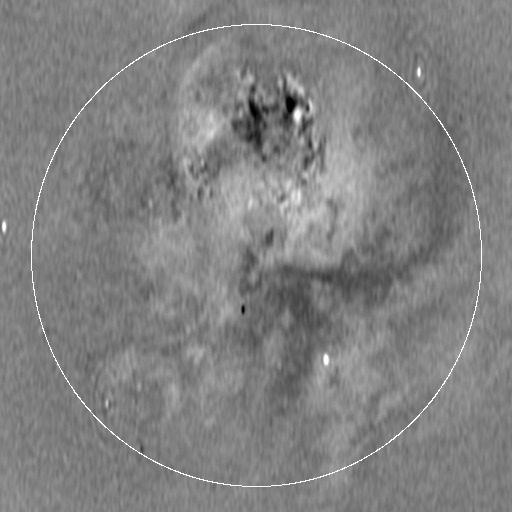}
  \addtocounter{figure}{1}\caption{Fig.~\ref{fig:nolc} (same grayscale) after image-based leakage
    correction.  The ``on-source'' correction of unresolved sources is accurate
    to 1\% of $I$, close to the theoretical precision of the measured leakage
    map, but the arcs around strong leakage remain unaffected.  Leakage
    amplitude differences between antennas produce rings, and phase differences
    produce asymmetric arcs. \label{fig:oldlc}}
  \end{minipage}
  
  \begin{minipage}[t]{1.0\columnwidth}
    \plotone{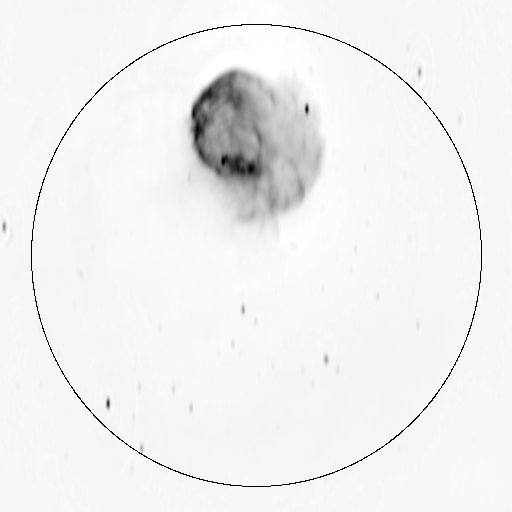}
    \addtocounter{figure}{-2}\caption{The sources of the $I$ radiation that leaked into
      Fig.~\ref{fig:nolc}.  The grayscale goes from -12 (white) to 350 (black)
      mJy/beam.}
  \end{minipage}\hfill\begin{minipage}[t]{1.0\columnwidth}
  \plotone{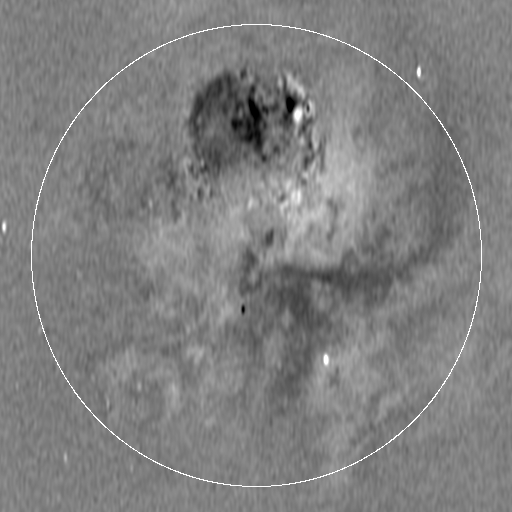}
  \addtocounter{figure}{1}\caption{Fig.~\ref{fig:nolc} (same grayscale) after correcting leakage using
    measured patterns for each antenna.  Leakage measurements were made only
    inside the circle, but they have been extrapolated to the edge of the
    image, which works well for clearing up the arcs of sources slightly
    outside the limit.  The remaining arcs are primarily due to differences
    between the primary voltage patterns of the antennas.  \label{fig:newlc}}
  \end{minipage}
\end{figure*}

If all of the antennas in an array are identical, the effects of the primary
beam and leakage patterns can be removed in the image plane.  At the Dominion
Radio Astrophysical Observatory (DRAO) we previously corrected the wide-field
contamination or polarization by multiplying the Stokes $I$ image with
``leakage maps'', and subtracting the results from the measured $Q$ and $U$
images, as in Fig.~\ref{fig:oldlc}.  The leakage maps were measured by
observing the apparent $Q/I$ and $U/I$ of an intrinsically unpolarized source
in a grid of offsets from the primary beam center \citep{bib:peracaula1999}.
This correction is performed completely in the image plane, so we call it the
``image-based'' leakage removal method.  It was immediately applicable for
DRAO's Synthesis Telescope (ST, \cite{bib:ST2000}) since its antennas are
equatorially mounted and thus its leakage patterns never rotate relative to the
sky.  The leakage patterns of an altitude-azimuth mounted telescope such as the
Very Large Array (VLA) rotate relative to the sky over the course of an
observation, but the image-based method can still be applied if the data are
first broken up into a series of snapshots \citep{bib:cotton_aips86}.

Unfortunately, there are differences, known or unknown, between the antennas of
any real interferometer.  The array may be a combination of antennas from
originally separate telescopes, such as the Combined Array for Research in
Millimeter-Wave Astronomy (CARMA, \citet{bib:carma2006}), and most very long
baseline interferometers.  It could also be in a transition period where only
some antennas have been modified, like the partially Enhanced Very Large Array,
and/or have serious surface errors as at (sub)mm wavelengths.  The ST is an
example of an array where the antennas are similar to each other, but with
known differences between them.  The two outermost antennas have 9.14\,m
diameters with four metal struts supporting their receivers, while the other
five are 8.53\,m in diameter with three struts, made of either metal or
fiberglass.  The differences in antenna diameter obviously create differences
in the half-power beamwidths (HPBWs), which at 1420\,MHz are 101.8\arcmin\ for
the two outer antennas and 108.8\arcmin\ for the rest.  The variation in the
number and composition of the struts affects the scattering of incoming light,
which is an important component of polarization leakage (most of the rest comes
from the feeds).

In polarization images the differences between antennas are seen as mismatches
between the standard point spread function (PSF, or ``dirty beam'') and the PSF
of the leakage.  When there are phase differences between the leakages of the
antennas, the effective PSF of the leakage is asymmetric \citep{bib:siss_ekers}
and offset from the peak of the unpolarized emission.  The effective PSF of the
leakage also varies across the field, meaning that subtracting a multiplication
of the Stokes $I$ map with a leakage map cannot fully correct the polarization
leakage of a heterogeneous array.  Additionally, the response of each antenna
in an array, both in leaked and true radiation, depends on the scale of the
source(s).  Resolved features have less power at high spatial frequencies, so
antennas that only participate in long baselines will contribute little leakage
to them.  Unresolved features have no such attenuation with baseline length,
and elicit an equally weighted mix of leakage from all antennas.  Usually
leakage maps are measured using a bright unresolved object, so in the case of a
heterogeneous array their corrections are only accurate for unresolved
sources.\footnote{Even then, only if the images are made with the same baseline
  weighting as used for the leakage map.}  Fig.~\ref{fig:oldlc} exhibits both
of these problems, as can be seen in comparison with Fig.~\ref{fig:newlc}, the
same image corrected with the method described in Section~\ref{sec:corr}.  The
main change for the unresolved sources is the presence or lack of surrounding
arcs, but the supernova remnant also shows a strong difference in the on-source
residual leakage.

The work described here aims to improve polarization imaging from the DRAO ST.
The telescope is engaged in an extensive survey of the major constituents of
the Interstellar Medium, the Canadian Galactic Plane Survey (CGPS,
\cite{bib:cgps2003}) which includes imaging in Stokes parameters $Q$ and $U$ at
1420 MHz along the plane of the Milky Way.

We recently measured the real and imaginary parts of the leakage patterns for
each antenna of the ST (similar to a hologram measurement, but with finer
spacing over a smaller area) and have started using them to correct its $Q$ and
$U$ observations, as seen in Fig.~\ref{fig:newlc}.

\section{Removing leakage from linear polarization for a heterogeneous array}
\label{sec:corr}

When the polarization leakage (or primary beam) varies with both direction and
baseline (i.e.\ antenna pair), there is no way to isolate their effects to one
of either the image or $uv$ planes.  Multiplying $V_{\mathrm{obs},AB}$ on the
left by $\jones{\Psi}{AB}^{-1}(\boldsymbol{n})$ does not work as it does
for \jones{D}{AB}, because $\boldsymbol{n}$ must be marginalized away
by integrating with $I_S$.  $I_S$ is the true intensity distribution of the
sky, which is unfortunately unknown.  A set of Stokes $I$ CLEAN
\citep{bib:hogbom74} components makes an acceptable substitute, however, both
in the replacement of the true sky by CLEAN components, and the temporary
neglect of $Q, U$, and $V$.

Since the correction for a heterogeneous array must be added directly to the
visibilities, the $I$ model used must match the true $I$ visibility function
within the sampled part of the $uv$ plane.  Specifically, it should not be
tapered by any sort of smoothing in the image plane, and the almost certain
discrepancies between the CLEAN components and true visibility function outside
the sampled part of the $uv$ plane are immaterial for this purpose.  The
CLEANed $I$ image should have small enough pixels to avoid quantization errors
in the component positions, and be CLEANed to at least a moderately faint
level.  Very faint $I$ emission does not need to be included since it will be
multiplied by the leakage, typically less than a few percent, and it tends to
have many more components, which would considerably slow down the calculation
of the correction.  Leakage from such emission could be quickly and adequately
removed by the image-based leakage map method, using the CLEAN residual image
as the $I$ map.  Calculating the correction for both $Q$ and $U$ of a CGPS
field, with a variable number, on the order of several thousand, of CLEAN
components, and $1.2 \times 10^5$ visibilities per polarization, takes from 15
minutes to overnight on a 2 GHz personal computer.

Assuming that $I_s = (I, 0, 0, 0)$ in correcting the wide-field leakage of
Eq.~\ref{eq:js} requires some care, since its validity depends on what Stokes
parameters are wanted, and whether the feeds are circularly or linearly
polarized.  In general each measured Stokes parameter is nominally the true
Stokes parameter, plus first order leakage from two of the other Stokes
parameters, plus second order leakage from the remaining one.  This comes from
the leakage Jones matrices for each antenna having only ones on diagonal, with
the leakage terms off-diagonal.  As a rule of thumb, the true $Q$ and $U$ can
be thought of as fractions of $I$, and $V$ as an even smaller fraction (i.e.\
second order).  With circularly polarized feeds the leakage of $I$ into $V$ is
second order, and thus possibly of the same magnitude as the leakage from
linear polarization, but the fact that $V = (RR - LL)/2$ means it is more
likely corrupted by errors in the right and left gains.  Linearly polarized
feeds replace $V$ with one of $Q$ or $U$ in a similar situation.  If necessary,
multiple Stokes parameters can be CLEANed to form an estimate of $I_S$, to be
iteratively improved using the procedure below.

The visibilities are corrected using the set of $I_S$ CLEAN components $V_C$ by
subtracting
\begin{equation}
  \label{eq:visleakage}
  {\cal L}_{\boldsymbol{b}_{AB}(t)} = \sum_{j} \jones{\Psi}{AB}(\boldsymbol{n}_j)
  V_{C,\boldsymbol{b}_{AB}(t),j}
\end{equation}
from the visibilities in each polarization at baseline
$\boldsymbol{b}_{AB}(t)$.  $V_{C,\boldsymbol{b}_{AB}(t),j}$ is the set of
visibilities in each polarization for antennas $A$ and $B$ at time $t$ for the
$j$th CLEAN component.  We prefer to use $V_{C,\boldsymbol{b}_{AB}(t),j}$ in
the form of Stokes parameters instead of feed correlations since usually only
one image ($I$) needs to be CLEANed before applying the correction.
\jones{\Psi}{AB} is therefore transformed into Stokes form, \jones{\Psi}{S,
  AB}:
\begin{equation}
  \label{eq:transf}
  \jones{\Psi}{S, AB} = \mathbf{\mathsf{S}}^{-1} \jones{\Psi}{AB}
  \mathbf{\mathsf{S}}.
\end{equation}

For the ST, with its circularly polarized feeds, the correction is only applied
to Stokes $Q$ and $U$ and second order leakages are ignored since the leakage
from $I$ to $Q$ and $U$ is first order.  That reduces the used portion of
\jones{\Psi}{S, AB} to linear combinations of elements of
\jones{\Psi}{A} and $\jones{\Psi}{B}^*$, allowing the leakages of
$I$ into $Q$ or $U$ for a given baseline to be easily calculated on the fly
from combinations of leakage maps for the individual antennas instead of
storing leakage maps for each combination of antennas:
\begin{equation}
  \label{eq:partleakages}
  l_{PAB} = l_{PA} + l_{PB}^*
\end{equation}
where $P$ is $Q$ or $U$.  Note that the imaginary part would be cancelled out
if $A$ and $B$ were identical.  The Jones matrices of individual antennas are
in circular coordinates ($p = R, q = L$), so
\begin{align}
  l_{QA} &= \phantom{-\mi(}\jones{\Psi}{A,12} - \jones{\Psi}{A,21}\textrm{, and}   \label{eq:lqu}\\
  l_{UA} &= -\mi\left( \jones{\Psi}{A,12} + \jones{\Psi}{A,21}\right).
\end{align}
These are the leakage patterns that are shown in Figs.~\ref{fig:qlvpre} to
\ref{fig:ulvpim}.  Note that the 12 and 21 subscripts refer to the off-diagonal
elements of the Jones matrix, not baselines between antennas 1 and 2.

\newlength{\lmswidth}
\ifPP
  \setlength{\lmswidth}{0.96\columnwidth}
\else
  \setlength{\lmswidth}{0.96\linewidth}
\fi

\begin{figure*}
  \begin{minipage}[t]{1.0\columnwidth}
    \ifOnline
    \includegraphics[width=\lmswidth,clip]{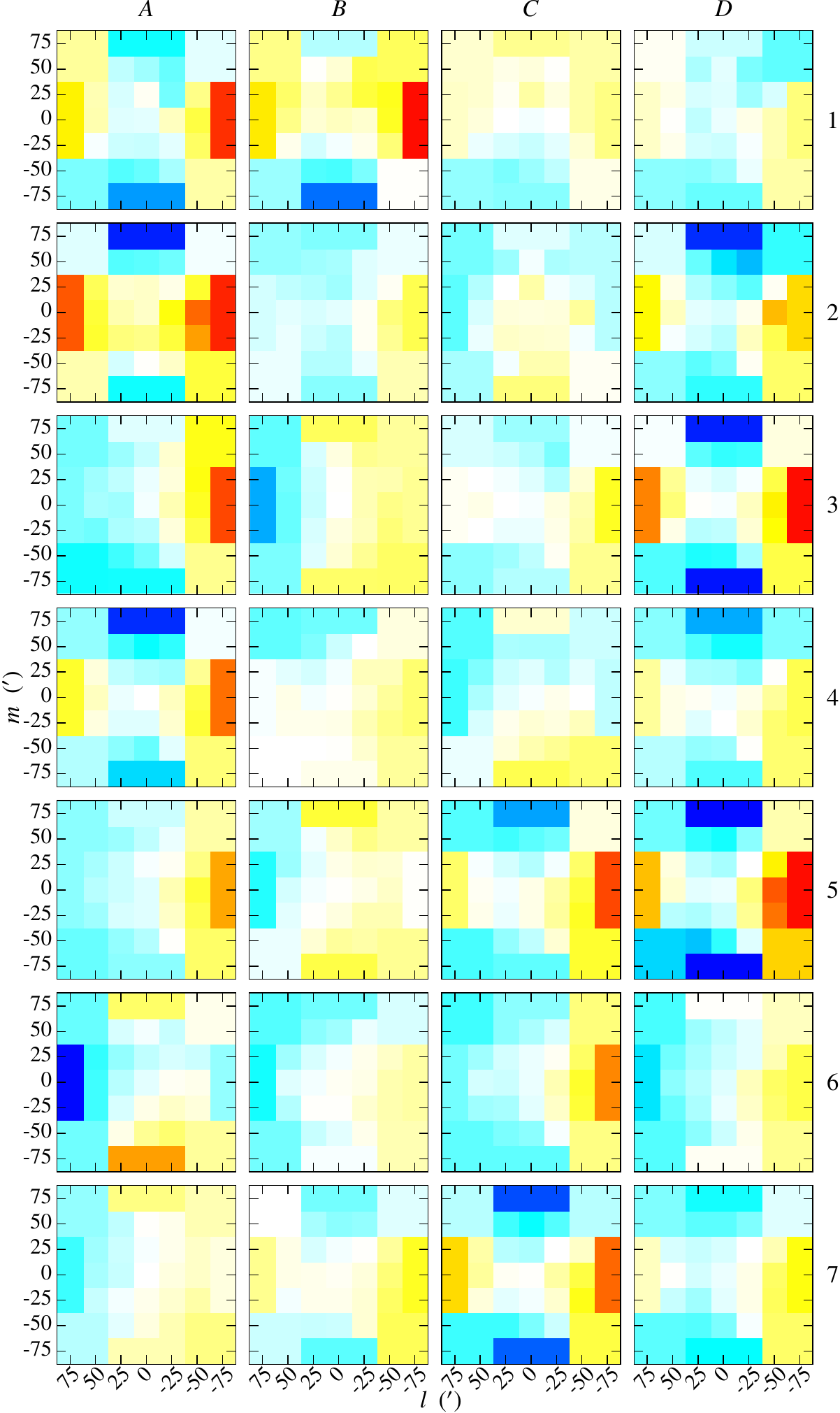}
    \else
    \includegraphics[width=\lmswidth,clip]{qlvpre_g}
    \fi
    \caption{Real parts of the voltage leakage from $I$ into $Q$ of antennas 1
      (top) to 7 (bottom) for bands A (left) to D (right).  The \rampname\ goes
      from -0.05 (\negcolor) to 0.05 (\poscolor).}
    \label{fig:qlvpre}
  \end{minipage}\hfill\begin{minipage}[t]{1.0\columnwidth}
    \ifOnline
    \includegraphics[width=\lmswidth,clip]{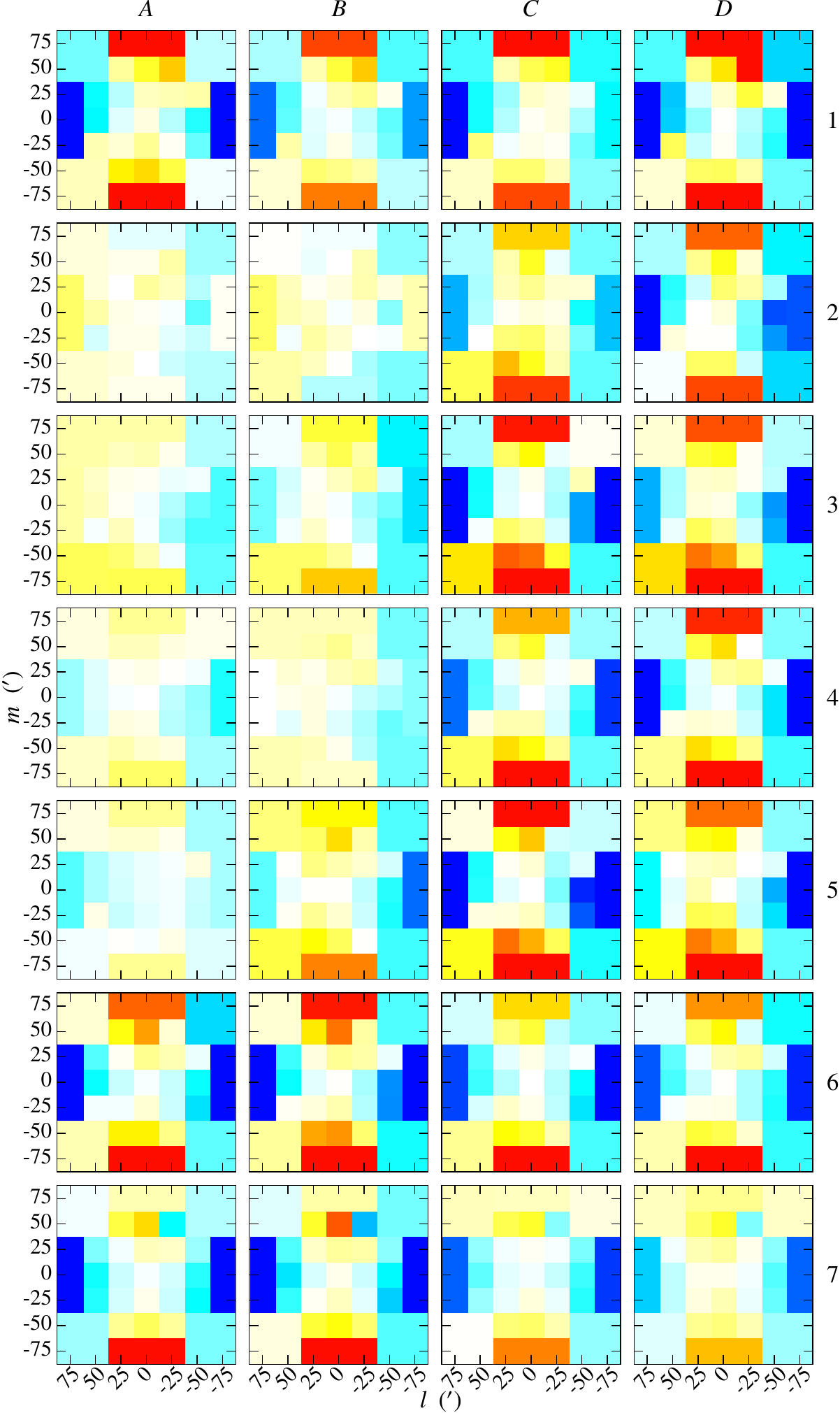}
    \else
    \includegraphics[width=\lmswidth,clip]{qlvpim_g}
    \fi
    \caption{Imaginary parts of the voltage leakage from $I$ into $Q$ of
      antennas 1 (top) to 7 (bottom) for bands A (left) to D (right).  The
      \rampname\ goes from -0.05 (\negcolor) to 0.05 (\poscolor).}
    \label{fig:qlvpim}
  \end{minipage}
\end{figure*}

\begin{figure}[tp]
\centering
  \ifOnline
    \includegraphics[width=\lmswidth,clip]{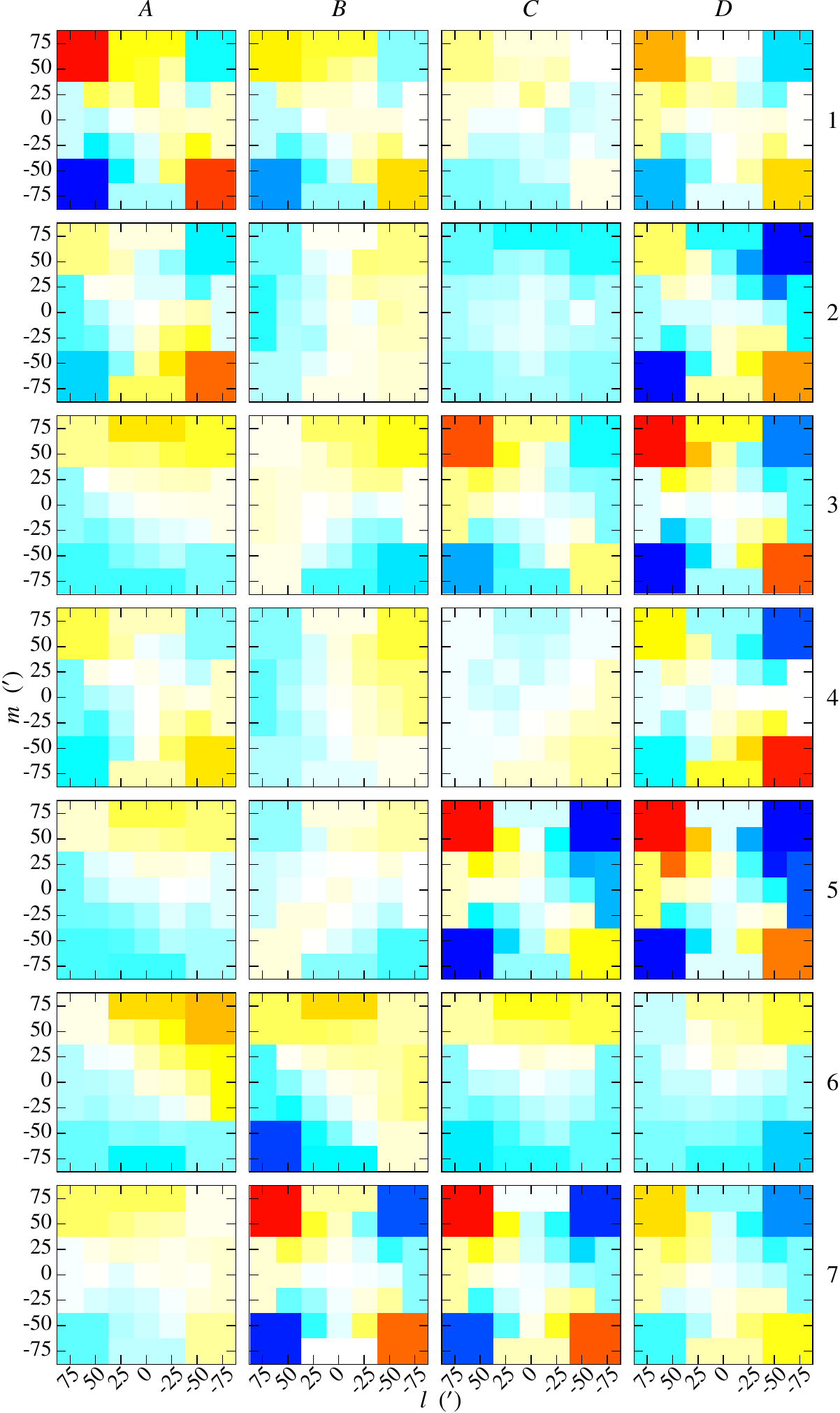}
  \else
    \includegraphics[width=\lmswidth,clip]{ulvpre_g}
  \fi
  \caption{Real parts of the voltage leakage from $I$ into $U$ of antennas 1
    (top) to 7 (bottom) for bands A (left) to D (right).  The \rampname\ goes
    from -0.05 (\negcolor) to 0.05 (\poscolor).}
  \label{fig:ulvpre}
\end{figure}

\begin{figure}[tp]
\centering
  \ifOnline
    \includegraphics[width=\lmswidth,clip]{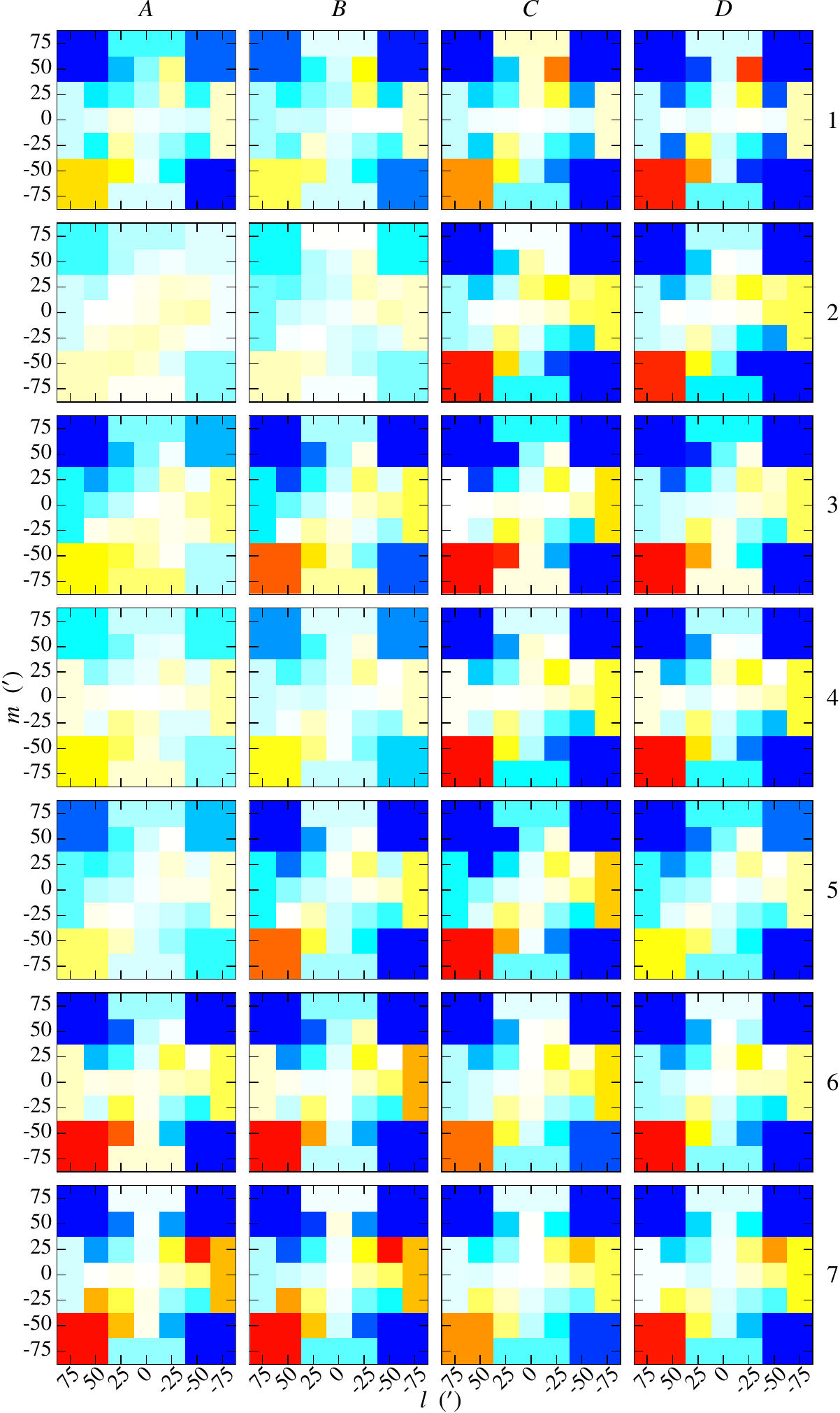}
  \else
    \includegraphics[width=\lmswidth,clip]{ulvpim_g}
  \fi
  \caption{Imaginary parts of the voltage leakage from $I$ into $U$ of antennas
    1 (top) to 7 (bottom) for bands A (left) to D (right).  The \rampname\ goes
    from -0.05 (\negcolor) to 0.05 (\poscolor).}
  \label{fig:ulvpim}
\end{figure}

\section{Simulated Leakage Maps}
\label{sec:simuleak}

\cite{bib:ng2005} calculated theoretical leakage voltage patterns for the ST's
three and four metal strut antennas.  Applying them to correcting polarization
leakage in the CGPS \citep{bib:cgps2003} confirmed that heterogeneity in the ST
was having a noticeable effect on the CGPS polarization images that was not
being corrected by subtracting the Stokes $I$ images multiplied by leakage
maps.  The correction still left significant residuals, however, which was not
surprising since the simulated patterns were based on an overly simplistic model
of the ST\@.  Some of the three-strut antennas have fiberglass supports for
their receivers.  Treating those as zero strut antennas would be incorrect
because each receiver box has cables running along one of its supporting
struts.  The unknown effective blockage of those cables, along with the partial
transparency of the fiberglass struts, made measuring the actual leakage
patterns essential.

\section{Antenna pattern measurements}
\label{sec:meas}

If one antenna, $A$, in an interferometer points directly at a bright isolated
source while the others look at it askew, $A$ will not have any off-axis
leakage or primary beam attenuation ($\jones{\Psi}{A}(\boldsymbol{0}) =
\jones{\Gamma}{A}(\boldsymbol{0}) = \mathbf{\mathsf{1}}$), and the
effective leakage and primary beam patterns will be those of the other antennas
alone.  Such offset observations with one antenna on axis are often done for
hologrammatic measurements of antenna surface errors, and with two
modifications the hologram scheme can be adapted to measure the leakage and
primary complex voltage patterns of each antenna.

The first modification is to compress the sampling grid of offsets.  Since
there is a Fourier transform relationship between the physical features of an
antenna and its angular power pattern, hologram measurements need to sample a
wide section of the celestial sphere to resolve small scale errors (i.e.\ a
misadjusted panel or smaller) on an antenna.  In an antenna pattern
measurement, however, it is more important to sample the main lobe well, so we
confined the sampling grid to within the first null.  In theory\footnote{Both
  the simulations of \citet{bib:ng2005} and the more intuitive realization that
  objects smaller than the antenna diameter, such as struts, produce features
  broader than the primary beam.} the antenna patterns should not vary any
faster with angle than the primary beam.  For the ST that means its patterns
should be fairly smooth on scales smaller than approximately a degree, so the
measurements were made on a grid with 25$^\prime$ spacing out to a maximum
distance of 75$^\prime$ from the beam center (the extent of beam used for
polarization mosaics).

The second modification is only in software, in that the antenna patterns come
directly from the measured visibilities, instead of requiring a Fourier
transform like surface error measurements.  The primary voltage pattern of an
antenna $B$ comes from a observation with an on-axis reference antenna $A$ of
an unpolarized and unresolved source $s$:
\begin{equation}
  \label{eq:hajs}
  V_{AB}^\textrm{obs} = \left( \jones{\Psi}{A}(\boldsymbol{0})
    \otimes \jones{\Psi}{B}^{*}(\boldsymbol{n})\right)
  \left( \jones{\Gamma}{A}(\boldsymbol{0})
    \otimes \jones{\Gamma}{B}^{*}(\boldsymbol{n})\right) \mathbf{\mathsf{S}}
  I_S.
\end{equation}
Since the source is effectively $I\delta(\boldsymbol{0})$ the integral of
Eq.~\ref{eq:js} was readily evaluated for Eq.~\ref{eq:hajs}.  It can be further
simplified by noting that $\jones{\Psi}{A}(\boldsymbol{0})$ and
$\jones{\Gamma}{A}(\boldsymbol{0})$ are identity matrices, and that
(unsurprisingly, given the physics it represents) the outer product has the
redistribution property (Eq.~5 of \citet{bib:hbs1}):
\begin{gather*}
  \left( \mathbf{\mathsf{M}}_A \otimes \mathbf{\mathsf{M}}_B \right)
  \left( \mathbf{\mathsf{N}}_A \otimes \mathbf{\mathsf{N}}_B \right) =
  \left( \mathbf{\mathsf{M}}_A \mathbf{\mathsf{N}}_A \right) \otimes
  \left( \mathbf{\mathsf{M}}_B \mathbf{\mathsf{N}}_B \right).
\end{gather*}
Eq.~\ref{eq:hajs} becomes:
\begin{align}
  V_{AB}^\textrm{obs} &= \left< v_s \otimes \left(
      \jones{\Psi}{B}(\boldsymbol{n})\, \jones{\Gamma}{B}(\boldsymbol{n}) v_s
    \right)^* \right>  \label{eq:shajs}\\
  &= \left[ \begin{array}{c}
      \phantom{-}\jones{\Gamma}{B,11}^{*}(\boldsymbol{n})
      \left< p_s^{ } p_s^* \right>
      + \jones{\Gamma}{B,22}^{*}(\boldsymbol{n})\,
      \jones{\Psi}{B,12}^{*}(\boldsymbol{n}) \left< p_s^{ } q_s^*
      \right>\\
      -\jones{\Gamma}{B,11}^{*}(\boldsymbol{n})\,
      \jones{\Psi}{B,21}^{*}(\boldsymbol{n}) \left< p_s^{ } p_s^* \right>
      + \jones{\Gamma}{B,22}^{*}(\boldsymbol{n}) \left< p_s^{ } q_s^* \right> \\
      \phantom{-}\jones{\Gamma}{B,11}^{*}(\boldsymbol{n}) \left< q_s^{ } p_s^*
      \right>
      + \jones{\Gamma}{B,22}^{*}(\boldsymbol{n})\,
      \jones{\Psi}{B,12}^{*}(\boldsymbol{n}) \left< q_s^{ } q_s^*
      \right>\\
      -\jones{\Gamma}{B,11}^{*}(\boldsymbol{n})\,
      \jones{\Psi}{B,21}^{*}(\boldsymbol{n}) \left< q_s^{ } p_s^* \right>
      + \jones{\Gamma}{B,22}^{*}(\boldsymbol{n}) \left< q_s^{ } q_s^*
      \right>      
    \end{array}\right] \nonumber
\end{align}
$v_s$ is $(p_s, q_s)$, the voltages that $s$ nominally imposes on the feeds.
$s$ is unpolarized, so $\left< p_s^{ } q_s^* \right> = \left< q_s^{ } p_s^*
\right> = 0$, and $\left< p_s^{ } p_s^* \right> = \left< q_s^{ } q_s^*
\right>$, reducing Eq.~\ref{eq:shajs} to
\begin{align*}
  V_{AB}^\textrm{obs} =  \frac{\left< p_s^{ } p_s^* \right> + \left< q_s^{ } q_s^*
      \right>}{2}\left[ \begin{array}{l}
      \phantom{-}\jones{\Gamma}{B,11}^{*}(\boldsymbol{n})\\
      -\jones{\Gamma}{B,11}^{*}(\boldsymbol{n})
      \,\jones{\Psi}{B,21}^{*}(\boldsymbol{n})\\
      \phantom{-}\jones{\Gamma}{B,22}^{*}(\boldsymbol{n})
      \,\jones{\Psi}{B,12}^{*}(\boldsymbol{n})\\
      \phantom{-}\jones{\Gamma}{B,22}^{*}(\boldsymbol{n})     
    \end{array}\right].
\end{align*}

The off-diagonal elements of $B$'s leakage Jones matrix are
\begin{align*}
  \jones{\Psi}{B,12}(\boldsymbol{n}) &= (V_{AB,qp}^\textrm{obs} /
  V_{AB,qq}^\textrm{obs})^*\textrm{, and} \\
  \jones{\Psi}{B,21}(\boldsymbol{n}) &= (V_{AB,pq}^\textrm{obs} /
  V_{AB,pp}^\textrm{obs})^*,
\end{align*}
which completely specifies \jones{\Psi}{B}, since the diagonal elements
are $1$.  

Measuring the primary voltage patterns requires knowing $\left< p_s^{ } p_s^*
\right> \quad \left( = \left< q_s^{ } q_s^* \right> \right)$.  Their diagonal
entries (the only nonzero ones) can be estimated\footnote{To within the noise,
  since the effects of the primary voltage patterns are defined to be whatever
  is left after on-axis calibration.} from a regular on-axis observation
$\left( \textrm{i.e.\ }\left <p_A^{ }(\boldsymbol{0})
    p_B^*(\boldsymbol{0})\right> \right)$, so
\begin{align*}
  \jones{\Gamma}{B,11}(\boldsymbol{n}) &\simeq \left( \frac{\left< p_A(\boldsymbol{0})
      p_B^*(\boldsymbol{n}) \right> }{\left< p_A(\boldsymbol{0})
      p_B^*(\boldsymbol{0}) \right> }\right)^*\textrm{, and} \\
  \jones{\Gamma}{B,22}(\boldsymbol{n}) &\simeq \left( \frac{\left< q_A(\boldsymbol{0})
      q_B^*(\boldsymbol{n}) \right> }{\left< q_A(\boldsymbol{0})
      q_B^*(\boldsymbol{0}) \right> }\right)^*.
\end{align*}

The 1420 MHz feeds of the ST are not offset from the central axes of the
antennas, so there should be no difference between its
$\jones{\Gamma}{B,11}(\boldsymbol{n})$ and
$\jones{\Gamma}{B,22}(\boldsymbol{n})$ because of beam squint.  We therefore 
collapse its primary voltage patterns from Jones matrices to a scalar for each
antenna:
\begin{gather*}
  g_{\textrm{off-axis}, B}(\boldsymbol{n}) \simeq \left[
    \frac{\left< p_A(\boldsymbol{0}) p_B^*(\boldsymbol{n}) \right>
    }{\left< p_A(\boldsymbol{0}) p_B^*(\boldsymbol{0}) \right> } +
    \frac{\left< q_A(\boldsymbol{0}) q_B^*(\boldsymbol{n}) \right>
    }{\left< q_A(\boldsymbol{0}) q_B^*(\boldsymbol{0}) \right> }\right] / 2.   
\end{gather*}
This approach can even be useful for telescopes with offset feeds, such as the
VLA, if care is taken to perform all calibration and self-calibration with $I =
(pp + qq)/2$ instead of $pp$ and/or $qq$ individually (conversation with J.
Uson, 2006).  In practice there is some error introduced for wide-field
polarimetry by approximating $\jones{\Gamma}{}$ with a scalar, since although
$\jones{\Gamma}{}$ does not mix polarizations in the observational basis, it
typically does in the Stokes basis.  For circularly polarized feeds squint
mixes $I$ and $V$ for directions away from the pointing center.  This does not
greatly contaminate $I$ since $V$ is almost always $\sim 0$, but is a serious
problem for measuring $V$, especially for continuum observations where
spectroscopic techniques cannot help.  The ST does have 1-2\% leakage from $I$
into $V$ at the half-power level of the primary beam, and although it could be
interpreted as squint the direction of the apparent squint sweeps through
180$^\circ$ as the frequency goes from band A to D.  The Robert Byrd Telescope
at Green Bank also sees a change in the direction of the apparent squint with
frequency \citep{bib:heiles2003}.  Such a variance with frequency is
inconsistent with the geometrical effect that affects the VLA.  The ST has only
been used to measure $V$ for exceptional cases like pulsars and the Sun, that
have strong circular polarization.  Observations that need to measure $V$
off-axis for more weakly polarized sources, especially in continuum, will need
to apply a more extensive treatment.  Similarly, when using linearly polarized
feeds \citep{bib:saultmemo} squint mixes $I$ with $Q$ instead of $V$, making
the $\jones{\Gamma}{11}(\boldsymbol{n}) = \jones{\Gamma}{22}(\boldsymbol{n})$
approximation less attractive.

Using $g$, the primary beam $B_{s, t}(\boldsymbol{n})$ for a baseline formed by
correlating antennas $s$ and $t$ is then
\begin{gather*}
  B_{s, t}(\boldsymbol{n}) = g_{\textrm{off-axis}, s}^{}(\boldsymbol{n})\,
  g_{\textrm{off-axis}, t}^*(\boldsymbol{n}).  
\end{gather*}
Note that the order of $s$ and $t$ matters when antennas $s$ and $t$ are not
identical.

Since the patterns are ratios, the requirement above that $s$ be unresolved can
be loosened to requiring that its size be much smaller than the angular scale
of variations in the primary beam, to avoid smearing the pattern samples.

With an interferometric array the patterns can be simultaneously measured for
all of the antennas except the reference antenna (i.e.\ $B$ is anything but $A$
in the above equations) by keeping only the reference antenna pointed at the
source while the other antennas look at it with the same grid of offsets.  The
patterns of the antenna used as a reference in that set of observations can be
measured by repeating the observations with a different antenna as the
reference.

\section{Observed Leakage Maps}
\label{sec:ob}

In order to minimize any effects from interference or crosstalk the antennas
were placed so that the distances between them were no smaller than 47\,m.
Observations were made of 3C 147, an unresolved bright source with a flux
density of 22\,Jy at 1420\,MHz.

The beams were sampled on a square grid with 25\arcmin\ spacing out to a
maximum radius of 75\arcmin\ from the beam center.  The time spent on each spot
was varied to achieve approximately the same uncertainty for each leakage
measurement, by making the integration intervals inversely proportional to the
nominal value of the primary beam:
\begin{gather}
  \label{eq:tint}
  t_{\textrm{int}}(\boldsymbol{n}) \propto \cos^{-6}\left( \frac{2\arccos(2^{-1/6})}{\textrm{HPBW}}|\boldsymbol{n}|\right).
\end{gather}
The on-axis pointing was observed longer because it was observationally
convenient and it is relatively important since it is used to normalize the
patterns.

The entire grid was observed twice, once with antenna 1 as the reference
antenna, and then again with antenna 7 as the reference antenna.  That allowed
the leakage maps and primary voltage patterns of all antennas in the ST to be
measured without requiring a separate reference antenna.

The leakage patterns were sampled out to 75\arcmin\ away from the beam center,
because that is the portion of the beam used by the CGPS.  Beyond that limit
(the 24\% power level of the beam) the leakages are expected to be large, and
require long integration times to measure with the same accuracy.  To help
remove errors that extend within the 75\arcmin\ from objects just outside it,
the leakage pattern measurements are extrapolated, using a nearest-neighbor
method, as far as 120\arcmin\ away from the beam center.  The leakage patterns
are also interpolated with cubic splines to a grid with 0.20\arcmin\ spacing to
match the pixels of the CLEAN component images.  An example correction with the
measured patterns of leakage from $I$ into $U$ is shown in Fig.~\ref{fig:newlc}.

\section{Quality of Leakage Correction}
\label{sec:results}

Since the form of primary beam used in Equation~\ref{eq:tint} is not
necessarily the correct one, the uncertainty in the primary voltage pattern for
antenna $A$, $g_{\textrm{off-axis},A}(\boldsymbol{n})$, is calculated as:
\ifPP
\begin{gather*}
  \left( \frac{\sigma_{g_{\textrm{off-axis},
          A}}(\boldsymbol{n})}{g_{\textrm{off-axis},
        A}(\boldsymbol{n})}\right)^2 = \frac{1}{n_{\textrm{samps},
      A}(\boldsymbol{n})} \left( \frac{\sigma_{I, 1 \textrm{samp},
        A}(\boldsymbol{n})}{I_A(\boldsymbol{n})}\right)^2 + \left(
    \frac{\sigma_{I_A}(\boldsymbol{0})}{I_A(\boldsymbol{0})} \right)^2,
\end{gather*}
\begin{gather*}
  \sigma_{g_{\textrm{off-axis}, A}}(\boldsymbol{n}) = \left(
    \frac{n_{\textrm{samps}, A}(\boldsymbol{0})}{n_{\textrm{samps},
        A}(\boldsymbol{n})} + \left| g_{\textrm{off-axis},
        A}\right|^2(\boldsymbol{n}) \right)^{1/2}
  \frac{\sigma_{I_A(\boldsymbol{0})}}{I_A(\boldsymbol{0})}.
\end{gather*}
\else
\begin{multline*}
  \left( \frac{\sigma_{g_{\textrm{off-axis},
          A}}(\boldsymbol{n})}{g_{\textrm{off-axis},
        A}(\boldsymbol{n})}\right)^2 = \\
  \frac{1}{n_{\textrm{samps}, A}(\boldsymbol{n})} \left( 
    \frac{\sigma_{I, 1 \textrm{samp},
        A}(\boldsymbol{n})}{I_A(\boldsymbol{n})}\right)^2 + \left(
    \frac{\sigma_{I_A}(\boldsymbol{0})}{I_A(\boldsymbol{0})} \right)^2,
\end{multline*}
\begin{multline*}
  \sigma_{g_{\textrm{off-axis}, A}}(\boldsymbol{n}) = \\
  \left( \frac{n_{\textrm{samps}, A}(\boldsymbol{0})}{n_{\textrm{samps},
        A}(\boldsymbol{n})} +
    \left| g_{\textrm{off-axis}, A}\right|^2(\boldsymbol{n}) \right)^{1/2}
  \frac{\sigma_{I_A(\boldsymbol{0})}}{I_A(\boldsymbol{0})}.
\end{multline*}
\fi
$g_{\textrm{off-axis},A}(\boldsymbol{n})$ gets its name from acting like
direction dependent factor of $A$'s gain.  $n_{\textrm{samps},
  A}(\boldsymbol{n})$ is the number of samples for antenna $A$ in direction
$\boldsymbol{n}$.

$l_{QA}(\boldsymbol{n})$ is calculated (for an antenna $A$ that comes before the
reference antenna, $B$) as
\begin{align*}
  l_{QA}(\boldsymbol{n}) &= \frac{1}{2}\left[ \frac{\left< R_A(\boldsymbol{n})
        L_B^*(\boldsymbol{0}) \right> }{\left< L_A(\boldsymbol{n})
        L_B^*(\boldsymbol{0}) \right> } + \frac{\left< L_A(\boldsymbol{n})
        R_B^*(\boldsymbol{0}) \right> }{\left< R_A(\boldsymbol{n})
        R_B^*(\boldsymbol{0}) \right> }
  \right]\\
  &= 
    \frac{\left< (R + \jones{\Psi}{A,12}L)(\boldsymbol{n})
        L(\boldsymbol{0}) \right> }{2\left< L(\boldsymbol{n})L^*(\boldsymbol{0})
      \right> } +
    \frac{\left< (L + \jones{\Psi}{A,21}R)(\boldsymbol{n})R^*(\boldsymbol{0})
      \right> }{2\left< R(\boldsymbol{n})R^*(\boldsymbol{0}) \right> }
    \\
    &= \frac{1}{2}\left[ \left( \frac{\left< R(\boldsymbol{n})L^*(\boldsymbol{0})
          \right> }{\left< L(\boldsymbol{n})L^*(\boldsymbol{0}) \right>
        }\right)_{nl} + \jones{\Psi}{A,12} + \left( \frac{\left<
            L(\boldsymbol{n})R^*(\boldsymbol{0}) \right> }{\left<
            R(\boldsymbol{n})R^*(\boldsymbol{0}) \right> }\right)_{\! nl} +
      \jones{\Psi}{A,21}\right].
\end{align*}
Note that the reference antenna is observing on-axis, so it has no leakage.
The uncertainty in $l_{QA}$ comes from the noise in the receivers:
\begin{gather*}
  \label{eq:sigmal}
  \left|\sigma_{l_{QA}}\right|^2 = \sum_{S = RR^*, LL^*, RL^*, LR^*} \left|
    \frac{\partial l_{QA}}{\partial S} \sigma_S \right|^2
\end{gather*}
The source is intrinsically unpolarized, so the crosscorrelations without
leakage, $\left< RL^*\right>_{nl}$ and $\left< LR^*\right>_{nl}$, are zero, and
thus so are the derivatives of $l_{QA}$ with respect to $RR^*$ and $LL^*$.  The
uncertainty of $l_{QA}$ reduces to
\begin{align*}
  \left|\sigma_{l_{QA}}(\boldsymbol{n})\right| &= \frac{2\sigma_Q}{\left<
      L(\boldsymbol{n})L^*(\boldsymbol{0}) \right> + \left<
      R(\boldsymbol{n})R^*(\boldsymbol{0}) \right> }\nonumber\\
  &= \frac{\sigma_Q}{|g_{\textrm{off-axis},A}(\boldsymbol{n})|I(\boldsymbol{0})}
\end{align*}
since antenna $A$ is the off-axis one.  $\sigma_{l_{UA}}$ has the same form,
and in our case is identical since $\sigma_Q = \sigma_U$.

The uncertainties are roughly independent of $\boldsymbol{n}$ because of the
time weighting, with an average value for antennas 2 to 6 of 0.0012.  The beam
centers are an exception, with average uncertainties for antennas 2 to 6 of
$6\times 10^{-4}$.  Antennas 1 and 7 were each used as reference antennas half
of the time, so their uncertainties are worse by a factor of nearly $\sqrt{2}$
(ameliorated by their slightly larger diameters).

\section{Discussion}
\label{sec:dis}

The measured leakage patterns, Figs.~\ref{fig:qlvpre} to \ref{fig:ulvpim}, show
that although there is some overall consistency in the patterns, their details
are unpredictable, both from antenna to antenna and from band to band in
frequency.  Most noticeably, the antennas with quadrupod receiver supports, 1
and 7, are structurally nearly identical, but their leakage patterns do not
show any more similarity to each other than they do to those of the tripod
antennas.  Likely this is because most of the leakage comes not from the
struts, but from the feeds.  The feeds are nominally identical, and their
individual flaws are neither easily apparent to visual inspection nor tied to
the type of antenna they are mounted on.  This suggests that wide-field
polarimetry with even nominally homogeneous arrays requires measuring the
leakage patterns of each antenna, if the needed fidelity warrants it.

Variation of the leakage patterns from band to band is prominent in
the real parts of the leakage patterns. This rapid change with 
frequency seems surprising at first glance: one might expect
properties of a waveguide feed to vary quite slowly with frequency, and
hardly at all across a band that is only 2\% of the center frequency. 
The cause appears to be the probes used to feed the reflector at 408
MHz; they are housed within the 1420 MHz feed \citep{bib:Veidt85}.
Computed simulations (B.G. Veidt, private communication) indicate that
these probes cause some fine structure in the performance at 1420
MHz.

The primary voltage patterns, Figs.~\ref{fig:pvpre} and \ref{fig:pvpim},
reassuringly exhibit only the expected dependence on wavelength; namely their
angular scales are proportional to the observing wavelength.  Their apparent
tight link to antenna structure suggests that primary voltage pattern errors
are more amenable to correction by adjusting the antennas, as is often done
using holograms.  Once the primary voltage patterns are known, their effect can
also be reduced post-observation, even for an inhomogeneous array
\citep{bib:bhatnagar_evla100}.  Currently such errors are attacked with
direction dependent self-calibration (modcal, \citep{bib:modcal}, also called
peeling), which is vulnerable to confusing true features on the sky with
unwanted artifacts.  Measuring the antenna patterns with a bright unresolved
calibration source instead of through self-calibration with a potentially
complicated fainter science target removes that vulnerability.

\begin{figure*}
  \begin{minipage}[t]{1.0\columnwidth}
    \ifOnline
    \includegraphics[width=\lmswidth,clip]{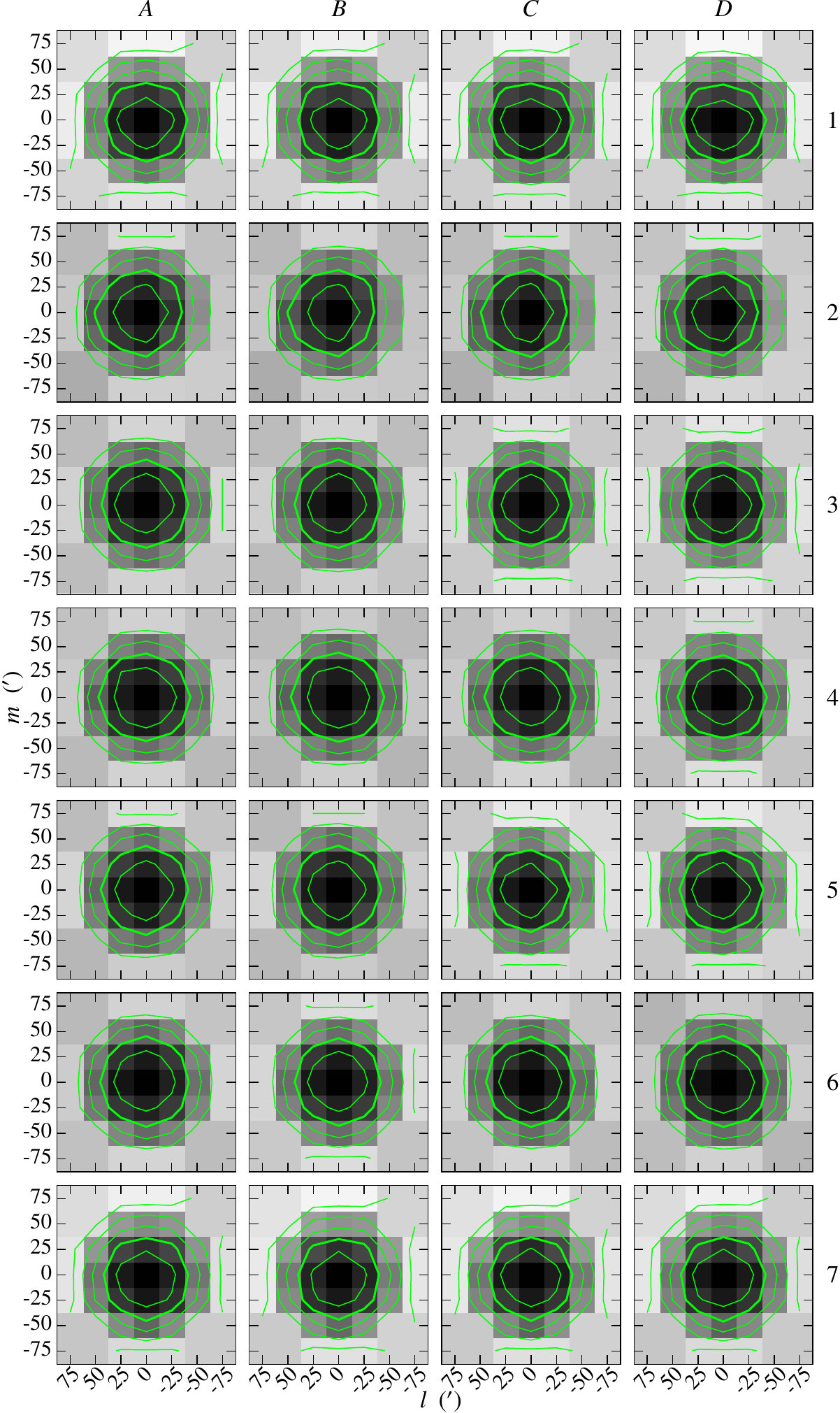}
    \else
    \includegraphics[width=\lmswidth,clip]{pvpre_g}
    \fi
    \caption{Real parts of the primary voltage patterns of antennas 1 (top) to
      7 (bottom) for bands A (left) to D (right).  The grayscale goes from 0.4
      (white) to 1.0 (black), and the contours go from 0.4 to 0.9 in steps of
      0.1.}
    \label{fig:pvpre}
  \end{minipage}\hfill\begin{minipage}[t]{1.0\columnwidth}
    \ifOnline
    \includegraphics[width=\lmswidth,clip]{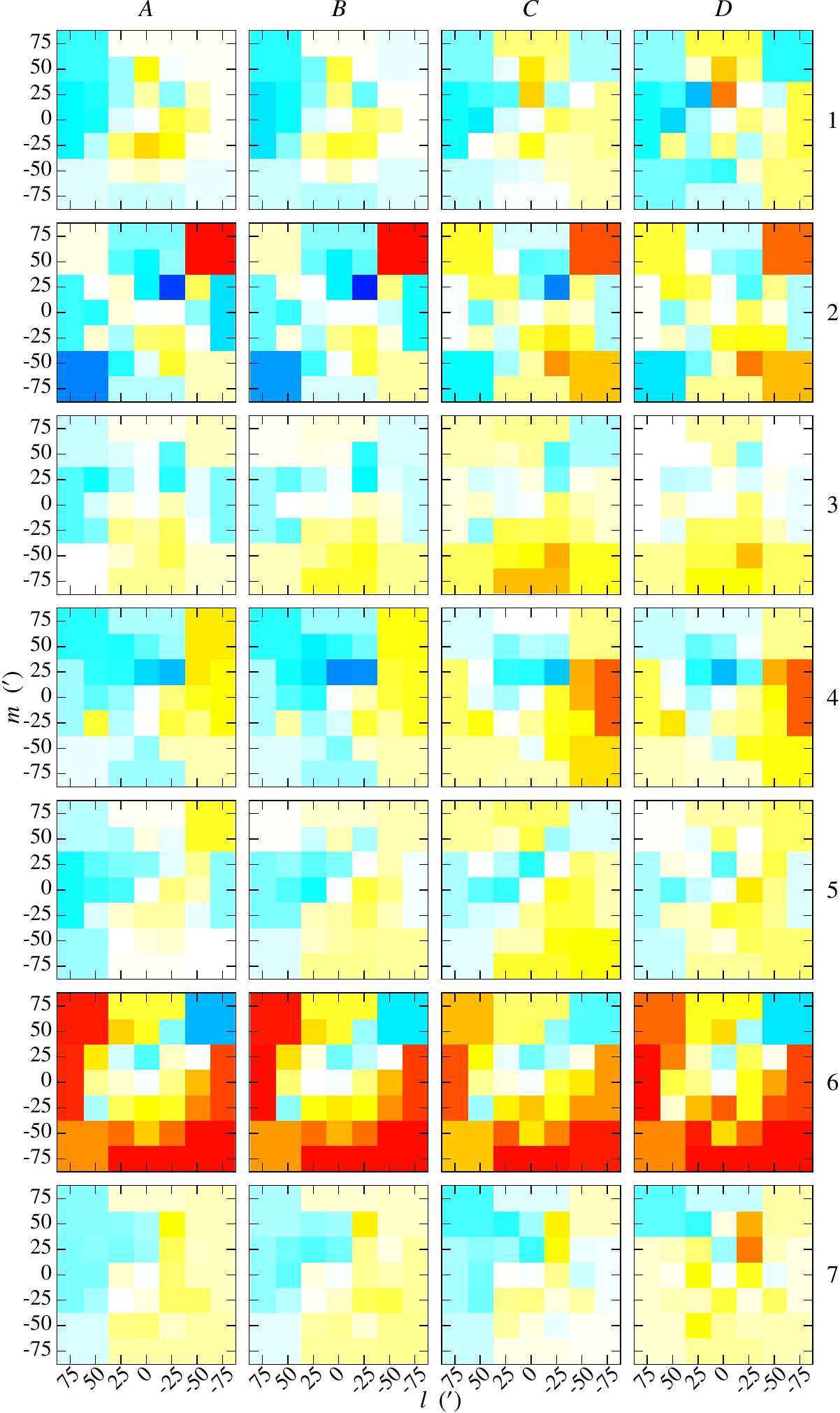}
    \else
    \includegraphics[width=\lmswidth,clip]{pvpim_g}
    \fi
    \caption{Imaginary parts of the primary voltage patterns of antennas 1 (top)
      to 7 (bottom) for bands A (left) to D (right).  The \rampname\ goes from
      -0.1 (\negcolor) to 0.1 (\poscolor).  The symmetry of antenna 6's patterns
      suggests that it is out of focus.}
    \label{fig:pvpim}
  \end{minipage}
\end{figure*}

Although we have only tested heterogeneous array leakage correction with
equatorially mounted antennas, in principle it would be even easier to adapt it
to antennas on altitude-azimuth mounts than the image-based leakage map method.
Since the leakage voltage pattern method already deals with visibilities on an
individual basis, the only modification needed would be make $x_j$ and $y_j$ in
Equation~\ref{eq:visleakage} functions of time to account for the rotation of
the antennas about the optical axis relative to the sky as the Earth turns.

An implicit, but difficult to avoid, assumption in correcting for the effect of
beam patterns is that the patterns do not change with time or observing
elevation.  The prospect of spending observing time on frequent antenna pattern
remeasurements, possibly for a set of elevations and frequencies, is
unappealing, so there is considerable pressure to engineer antennas that are
stable enough for occasional measurements to capture most of the effects.  The
ST antennas were not expected to change significantly with time or observing
direction, but we confirmed their behavior by comparing recent leakage
measurements to the measurements made by Peracaula of the ST's overall leakage
amplitude maps at 21 cm wavelength.  There was little change over the
intervening 10 years, despite some surface modifications to a few of the
antennas.  Stability is expected to be a more serious problem for larger
(as measured in wavelengths) dishes, especially if standing waves create a
noticeable resonance effect in the leakage at certain observing frequencies.
Interpolation, or theoretical modeling, may be useful for extending the
applicability of measured maps to additional elevations and/or frequencies.
Alternatively, if an extremely accurate correction is only needed for one
bright source within the field of an observation, the antenna patterns could be
measured at that spot immediately before and after the science observation, as
opposed to mapping the entire main lobe of the antenna patterns.

\acknowledgements

The National Radio Astronomy Observatory is a facility of the National
Science Foundation operated under cooperative agreement by Associated
Universities, Incorporated.  The Dominion Radio Astrophysical Observatory
is operated as a national facility by the National Research Council of
Canada.  The Canadian Galactic Plane Survey is a Canadian project with
international partners.  The survey is supported by a grant from the
Natural Sciences and Engineering Council (NSERC).  We thank the reviewers
for their time and helpful comments.

R. Kothes kindly pointed out IC443 as a good example of the new correction
method's efficacy, and D. Routledge helpfully expanded upon the simulations
of Ng et al.  We appreciate the assistance of J.E. Sheehan in facilitating
the measurements and D. Del Rizzo in partially processing the data.  The
processing was also assisted by K. Douglas' programming and documentation
for antenna surface measurements.  R.R. appreciates J.~Uson, W.~Cotton,
C.~Brogan, and D.~Balser sharing their experience with the VLA and GBT.

 
\bibliographystyle{apj}
\bibliography{plp}

\end{document}